\documentclass{aa}

\usepackage{graphicx}
\usepackage{txfonts}

\begin{document}

\title{Quasars and Galaxy Clusters Paired Across NGC 4410}

\author{H. Arp
       \inst{1},
        E.M. Burbidge
       \inst{2}
       \and
        D. Carosati
       \inst{3}
       }

\offprints{H. Arp}
        
       \institute{Max-Planck-Institut f\"ur Astrophysik, Garching
                  85741, Germany\\
                  \email{arp@mpa-garching.mpg.de}
      \and
         Center for Astrophysics and Space Sciences 0424,
        University of California San Diego, CA 92093-0424
    \and  
        Armenzano Astronomical Observatory, 06081 Assisi (PG), Italy}

\date{Received ....; Accepted ...}

\abstract 
{}
{The aim is to investigate the region of the sky around
NGC4410/Mrk1325 for objects which are physically associated with this
active, double nucleus galaxy.}
{We use archived data to study the placement, brightness, X-ray
properties and redshifts of objects within 60' of the bright, central galaxy.}
{It is found that pairs of quasars are aligned across NGC 4410
which, if ejected from it, have equal and opposite ejection velocities
and fall very close to the quantized Karlsson redshift peaks for
quasars. X-ray sources and Abell galaxy clusters at higher
redshifts appear elongated along directions away from NGC4410.}
{}



\titlerunning{Associations with NGC4410} 

\maketitle

\section{Introduction}

NGC 4410 is a disturbed galaxy classified as Sab interacting. It
appears to contain two nuclei, 4410 A and B, separated by about 
20 arcsec. A is more compact, an X-ray source and probably the source
of the ultraviolet continuum which caused NGC 4410 to be catalogued as
Markarian 1325 (Markarian et al. 1980). 

Radio measures by Beverly J. Smith (2000) show that A
has abundant molecular gas. There is HI coincident with faint
optical tails to the SE and NE. To the NE there is a longer 
X-ray structure and optical bridge along which lie a second and third
galaxy. Optical measures (B. Smith et al.2003) indicate a redshift of
cz = 7440 km/sec for A and 7500 for B. The HI redshift for A + B is
listed as 7350 km/sec.

The above reference shows there are 8 point x-ray sources around the
active Markarian nucleus, A. Because they are in the Ultra Luminous
class (ULX), it has been speculated that they are black hole
binaries or background sources. There has been no attempt to measure
their redshifts nor any reason advanced as to why they are closely
situated around the component which is the Markarian galaxy. X-ray
observations are discussed by (Tsch\"oke, et al 1999) among others.

\section {Quasar Pairs Associated With NGC 4410}

Among the closest objects to NGC 4410 there are two optically bright
quasars at 17.5' and 23.6' distance which are very exactly aligned
NE-SW across the galaxy center. They are the closest pair shown in
Fig. 1. The exact alignment and close centering of these two inner
quasars across the active galaxy nucleus is strong evidence for them
to be associated with NGC 4410. Moreover, X-ray observations of this
region reveal that both of these quasars are exceptionally bright and
similar at C = 23.8 and 27.9 counts/kilosec (ROSAT PSPC).

Quasar redshifts have long been found to favor certain periodic values
of which the pairs in Table 1 are an example (Karlsson 1971; Arp et
al. 1990):

$$Karlsson Periodicities:  .06, .30, .60, .96, 1.41, 1.96, 2.64,
. . .$$
 
We have to note, however, that, even when their redshifts are
transformed to the frame of the central galaxy ($z_0$)'s, that the
innermost pair
of quasars $z_0$'s are not close to a Karlsson, preferred redshift
peak. What is the reason for this deviation? {\it It turns out that
when we average the two $z_0$'s we obtain $|z_0| = .59$ which is
almost exactly the expected intrinsic redshift peak of $z_K = .60$.}
We then see that the individual deviations from the preferred  peak
are $z_v$ = -.064 and +.050. In other words the ejection velocity is
away from us for one of the pair and towards us for the other in the pair. 

Another trio of catalogued quasars, somewhat more separated but closely 
matching each other in redshift are located NW and SE of NGC
4410. They are shown in Fig. 1. In Table 1 these quasars are shown
to have mean $|z_0|$'s close to the Karlsson peak $z_K$ =
1.41. These quasars are apparently ejected more across the line of sight
because they are more separated and have smaller projected components
of ejection velocity (as computed in table 2).

\begin{table}[ht]
\caption{Pairs Across NGC 4410} \label{Table1} \vspace{0.3cm}
\begin{tabular}{lccclc}
Quasar & d' & mag. & z & $z_0$ & $<z_0>$\\
& & \\

2E 1224+0930   & 17.5' & 18.5g & .722  & .681\\
               &       &       &       & mean & .590\\    
LBQS 1222+0901 & 23.6  & 17.3  & .535  & .498\\
                                            \\
                                            \\
LBQS 1222+0928 & 25.4  & 18.5g & 1.466 & 1.407\\
               &       &       &       & mean & 1.410\\
LBQS 1225+0836 & 49.1  & 17.83 & 1.471 & 1.412\\
                                              \\     
                                              \\
SDSS 1225+0955 & 58.6  & 18.8g  & 1.429 & 1.371\\

\end{tabular}
\end{table}

\section{All Catalogued Quasars within 60' of NGC 4410}

The disposition of all catalogued quasars within a degree of the
central galaxy are listed in Table 2. On studying these 25 quasars
within 60' of the central galaxy it becomes apparent that the quasars
with the smallest residual ($z_v$) \footnote {$z_v$ is the residual
from the Karlsson value which can be interpreted as a peculiar
velocity: (1 + $z_0$)/(1 + $z_K$) = (1 + $z_v$)} from periodic $z_0$
values involve
brighter quasars and are generally at greater distances from NGC
4410. This is precedented since paper III of ``The 2dF Redshift
Survey'' (Arp and Fulton 2006) showed shells of quasars vacant for
radii of 10' - 20' around some active parent galaxies. This was
viewed as ejected quasars hitting a boundary shell, slowing and
completing their evolution at some distance from the parent galaxy.  

In any case NGC 4410 is a large bright galaxy and one would expect its 
quasars to be brighter and at greater separation  than from
smaller, fainter parents that happen to be in the area. In order to
make this a quantitative statement we plot in Fig. 2 the frequency
distribution of $z_v$ for radii greater than 30'. One sees the
characteristic double peak around $z_v$ = 0 for quasars with small
positive and negative ejection velocities. The open circles represent
quasars within 30' radius and are rather evenly spread throughout the
$z_v$ range.

In order to establish  that the small $z_v$'s systematically belong
to the bright NGC 4410 we show Fig. 3 where they average about 0.7
mag. brighter than quasars with  $z_v \geq .03$

\begin{table}[ht]
\caption{Quasars Near NGC 4410} \label{Table2} \vspace{0.3cm}
\begin{tabular}{lcccl}
Object & $z$ & $z_v$ & d' & mag.\\
& & \\

         NGC 4410 & .0244 & ---- & 0.0 &  13.6\\
         SDSS    & 2.237 & +.067 &  9.6 & 19.4g\\
         SDSS    &  .622 & +.014 & 13.3 & 18.9g\\
         SDSS    & 1.903 & -.043 & 14.4 & 19.3g\\
         HB89    &  .722 & +.050 & 17.5 & 18.5g\\
         LBQS    &  .535 & -.064 & 23.6 &  17.3\\
         SDSS    & 1.776 & -.085 & 25.2 &  19.1g\\
         LBQS    & 1.466 & -.002 & 25.4 &  18.5g\\
         HB89    &  .084 & -.007 & 26.9 &  16.8\\
         SDSS    & 1.590 & +.049 & 29.0 &  20.0g\\
         SDSS    & 1.502 & +.013 & 30.9 &  18.6g\\
         SDSS    &  .628 & -.007 & 34.8 &  19.0g\\
         SDSS    & 1.043 & +.017 & 39.5 &  18.5g\\
         SDSS    & 1.363 & -.043 & 43.0 &  19.1g\\
          PSS    & 4.340 & ---   & 48.9 &   ----\\
         LBQS    & 1.471 &  .000 & 49.1 &  17.83\\
         SDSS    & 1.076 & +.033 & 49.7 &  18.7g\\
         SDSS    & 1.345 & -.051 & 49.8 &  19.1g\\
         SDSS    & 1.090 & +.040 & 50.3 &  19.0g\\
         SDSS    & 1.715 & +.100 & 51.3 &  19.1g\\
         LBQS    &  .681 & +.025 & 52.0 &  17.6\\
         SDSS    & 2.649 & -.022 & 54.2 &  19.9g\\
         SDSS    &  .773 & +.081 & 56.3 &  18.5g\\
        2MASX    &  .064 & -.025 & 56.3 &  16.7g\\
         LBQS    &  .397 & +.049 & 56.4 &  18.2g\\
         SDSS    & 1.429 & -.017 & 58.6 &  18.8g\\

\end{tabular}
\end{table}

\section{X-ray Sources Related to NGC 4410}

We have seen two strong X-ray sources aligned NE - SW across the
central, active galaxy. Are there more X-ray sourcess in the vicinity? 
Fig. 4 shows the fainter X-ray sources in a 1 degree radius around
NGC 4410 as catalogued in the ROSAT PSPC Source Browser. It is
apparent that there is a strong cluster about 18' SE from the galaxy.

Within $\sim 15^o$ radius directly around NGC 4410 there is a
relative vacancy of sources (similar to cases referenced earlier in
Paper III of Arp and Fulton 2006). But most important of all, Fig. 4
shows, from the thickest part of the bounding arc, a dense
concentration of points streaming away to the SE from just the
direction which leads back to the galaxy. That elongation of X-ray
sources is mostly due to galaxies in a cluster. Not just diffuse X-rays
within a cluster volume but in this case probably individual X-ray
sources of objects within a cluster.

\subsection{Abell Clusters 1541, 1541A and 1541C}

The most surprising result now comes from the measures of the
redshifts of  galaxies in this region  {\it It turns out there are
three clusters, all listed at exactly the same position in this
concentration of X-ray sources:}

              $$ Abell~~ 1541~~~~ z = .089$$                

              $$ Abell~~ 1541A~~~ z = .0244$$

              $$ Abell~~ 1541C~~~ z = .0035$$ 

The redshift z = .0244 is exactly the redshift of NGC 4410. So the
companion galaxies of NGC 4410 are shown to be spatially contiguous
with the much higher redshift cluster Abell 1541 at z = .089.

Supporting evidence that the  z = .089 cluster Abell 1541 was
ejected from the NGC 4410/Mrk 1325 active galaxy is to be found in
Fig. 5.  Here we have plotted all galaxies catalogued in NED within
$\sim$ 60' of NGC 4410. {\it It turns out that about 26' NW from NGC
4410 is an elongated group of galaxies with $z_{ave} = .091$. They are
strikingly paired with the Abell custer 1541 at z = .089 on the
opposite side of NGC 4410.} This is the canonical pattern of elongated
clusters paired on either side of an active galaxy.

The physical nature of the pairing is further supported by Fig. 6
where the X-ray sources of Fig. 4 are now superposed onto the galaxy
distribution of Fig. 5. One can better see the circular boundary from
East to South of sources around NGC 4410. One can see evidence of
connection NW to the z = 0.91 elongated cluster of galaxies. But most
strongly of all, the increasing density and elongation of sources to
the SE from NGC 4410. The ejection argument for X-ray sources would be 
hard to argue against.

\section{Previous Associations of Higher Redshift Galaxy Clusters with
Active Galaxies}

It should be emphasized that the NGC 4410 pairing is closely the same
result that was obtained for the Abell X-ray clusters Abell 3667 and
3651 (Arp and Russell 2001). Abell 3667 was later shown to be moving
away from its lower redshift parent at a speed of 1400km/sec (Arp 2003 pp 178
-186). Other examples were shown in the 2003 reference including NGC
7131 with an elongated cluster of z = .088 pointing back to the z =
.018 parent (page 194). In another case, from Arp 220 (z = .018), there
emerges a group of X-ray galaxies at z = .09 (Arp et al. 2001, Arp
2003). An elongated X-ray cluster showed near perfect pairing with the
X-ray ejecting NGC 720 (Arp 2005, Burbidge et al 2006). As shown here in
Fig. 7 the Arp 220 case is a particularly clear demonstration of the
NGC 4410 ejection interpretation because the X-ray group is right
along the line of the quasar pair All the redshifts are strikingly
similar between NGC 4410 and Arp 220.

It should also be noted that Abell 1541, here associated with NGC
4410, is a strong X-ray cluster. In the SHARC Survey for X-ray
clusters it was one of the 37 brightest detected (Romer et al. 2000). 
Among bright X-ray cluster it is noted that they tend to occur in
elongated, non-equilibrium forms. In addition to the ones cited above
there is the elongated X-ray cluster which appears like a jet coming out
of the Seyfert Galaxy NGC 5548 (Arp 1998 p 145).

\section{The Origin of NGC 4410}

Considering the active nature of NGC4410/Mrk1325 discussed in the
Introduction it is natural to ask: ''Where is its lower redshift
parent? '' Actually the answer is immediately present in the Abell
Clusters catalogued SE of NGC 4410. Abell 1541C has z = .0035 and
contains large galaxies any one of which could be the parent of NGC
4410. The redshift z = .0035 is a little over 1000 km/sec and, in
position and redshift, places these galaxies in the conventional
Virgo Cluster. Within a cluster like Virgo the interglactic medium
would presumably furnish a resisting force which would string out the
clusters in the direction of ejection. The NVSS radio map presently
suggests that there is roughly a ring of radio sources around NGC 4410 
which represents the major onset of interaction of the ejecta with the
medium.      
 
The NGC 4410 galaxies at z = .0244 (cz = 7320 km/sec) are then part of
that over density of higher redshift galaxies in the Virgo direction
which are conventionally believed to form a background cluster.(see
Arp 1998 p 69). But they are like Stephan's Quintet at 5700 - 6700 km/sec
which is connected to NGC 7331 (Arp 1987 p 99) and the Cartwheel
galaxy and companions which are associated with NGC 134. Also the
6400 - 6700 km/sec companions to NGC 4151 (Seeing Red, page 78).

A final comment on periodicity is to note that the Abell 1541 cluster 
at z = .089 when referenced to its parent NGC 4410 has $z_0$ = .063.
The lowest Karlsson period is $z_K \sim .06$ to .065 for active galaxies
approaching AGN/quasar properties. (see G. and E.M. Burbidge 1967 and Arp et
al. 1990 for the discovery of this lowest periodicity). Of course this
periodicity would apply to the associations of z = .09 galaxies with z
= .02 active parents mentioned in the last paragraph of the previous
section. As for secondary ejections, two of the faintest quasars in
Table 2 (19.4g and 20.0g) fall $z_v$ = +.004 and -.013 from the z =
.089 cluster redshift.       

\section{Summary and Conclusions}

Investigation of the field around the very active galaxy
NGC4410/Mrk1325 reveals pairings of quasars and higher redshift
galaxies including clusters of galaxies. The patterns are supported in
detail by many previous investigations. The fit of the redshifts with 
the long standing Karlsson periodicity relation confirms again the
reality of the numerical values of that series and at the same time the
reality of the physical associations of the higher redshift objects. 

The suggested model is that high redshift quasars are ejected and
evolve in steps to lower redshifts. Either initially or as they grow
in luminosity and mass they can divide into smaller objects which can
evolve into groups or clusters of smaller companion galaxies. The
elongated distribution of clusters implies that the evolving low mass
plasmoids can be disrupted by passing out through the intergalactic
medium which is often observed to be blown partially clear
around the ejecting parent.

\section*{References}

\noindent Arp, H. 1987, Quasars, Redshifts and Controversies,
Interstellar Media, Berkely
\medskip

\noindent Arp, H. 1998, Seeing Red: Redshifts, Cosmology and Academic
Science, Apeiron, Montreal
\medskip

\noindent Arp, H. 2003, Catalog of Discordant Redshifts, Apeiron, Montreal 
\medskip

\noindent Arp, H.  2005 arXiv:astro-ph/0510173
\medskip

\noindent Arp, H., Bi, H., Chu, Y., Zhu, X. 1990, A \& A 239, 33
\medskip

\noindent Arp, H. and Russell 2001, ApJ 549, 802
\medskip

\noindent Arp, H., Burbidge, E. M., Chu, Y., Zhu, X. 2001, ApJ 553
\medskip

\noindent Arp, H., Fulton, C. 2006, The 2dF Redsfuft Survey, Paper
III, to be submitted
\medskip

\noindent Burbidge, G. and Burbidge E. M., 1967 ApJ 148, L107
\medskip

\noindent Burbidge, E. M., Guti\'errez, C., Arp, H. 2006, PASP 118, 124 
\medskip

\noindent Karlsson, K. 1971 A\&A 13, 333
\medskip

\noindent Markarian B., Lipovetskij V., Stepanian J., Astrophysics,
1980, 15, 363-376 
\medskip

\noindent Romer, A., Nichol, R., Holden, B. et al. 2000, ApJS 126, 209
\medskip

\noindent Smith, B. 2000, ApJ 541, 624, 
\medskip

\noindent Smith, B., Nowak, M., Donahue,, Stocke, J. 2003, AJ 126,
1763  
\medskip

\noindent Tsch\"oke, D., Hensler G., Junkes N., 1999, A\&A 343, 373

\clearpage

\begin{figure}
\includegraphics[width=8cm]{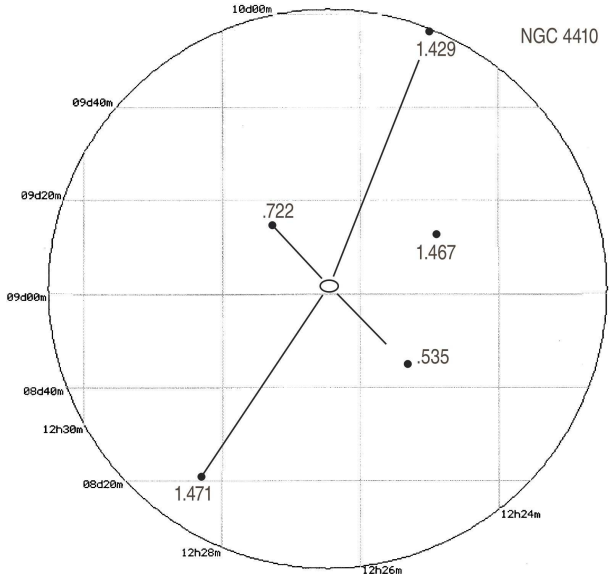}
\caption{The most conspicuous pairs of quasars within 60' of the
active galaxy NGC4410/Mrk1325 are shown. See Tables 1 and 2
\label{fig1}}
\end{figure}

\medskip
\medskip

\begin{figure}[p]
\includegraphics[width=9cm]{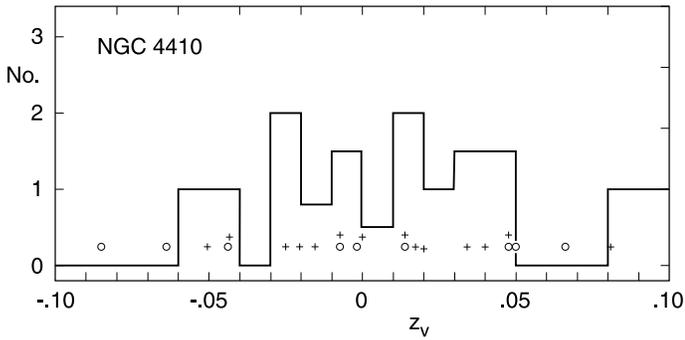}
\caption{The residuals from the Karlsson redshift periodicities,
$z_v$, are  $d > 30'$ pluses, and $< 30'$ circles. histogram is for +'s.
\label{fig2}}
\end{figure}

\begin{figure}
\includegraphics[width=6cm]{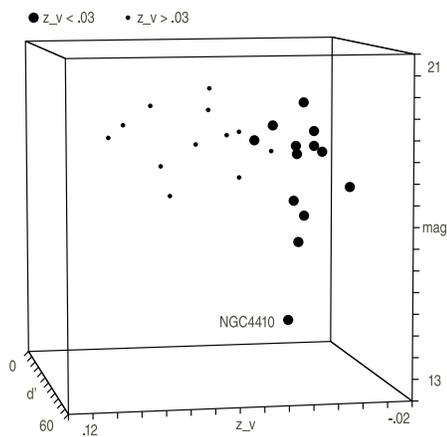}
\caption{Plot showing that periodicity residuals $z_v \leq .03$ average
brighter in apparent magnitude than $z_v \geq .03$
\label{fig3}}
\end{figure}

\begin{figure}
\includegraphics[width=8cm]{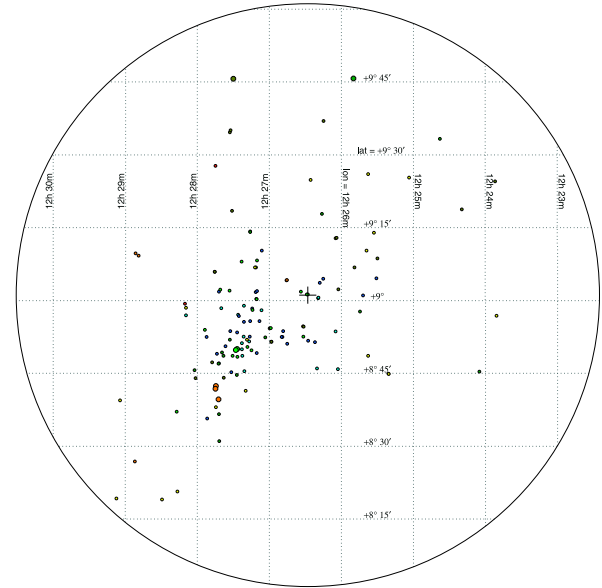}
\caption{The dense elongation of X-ray sources SE of NGC 4410 (plus sign at
center) coincides with the Abell Clusters A1541, A 1541A and A1541C
(12h27m27s+08d50m24s). From Rosat source browser.
\label{fig4}}
\end{figure}

\clearpage

\begin{figure}
\includegraphics[width=11cm]{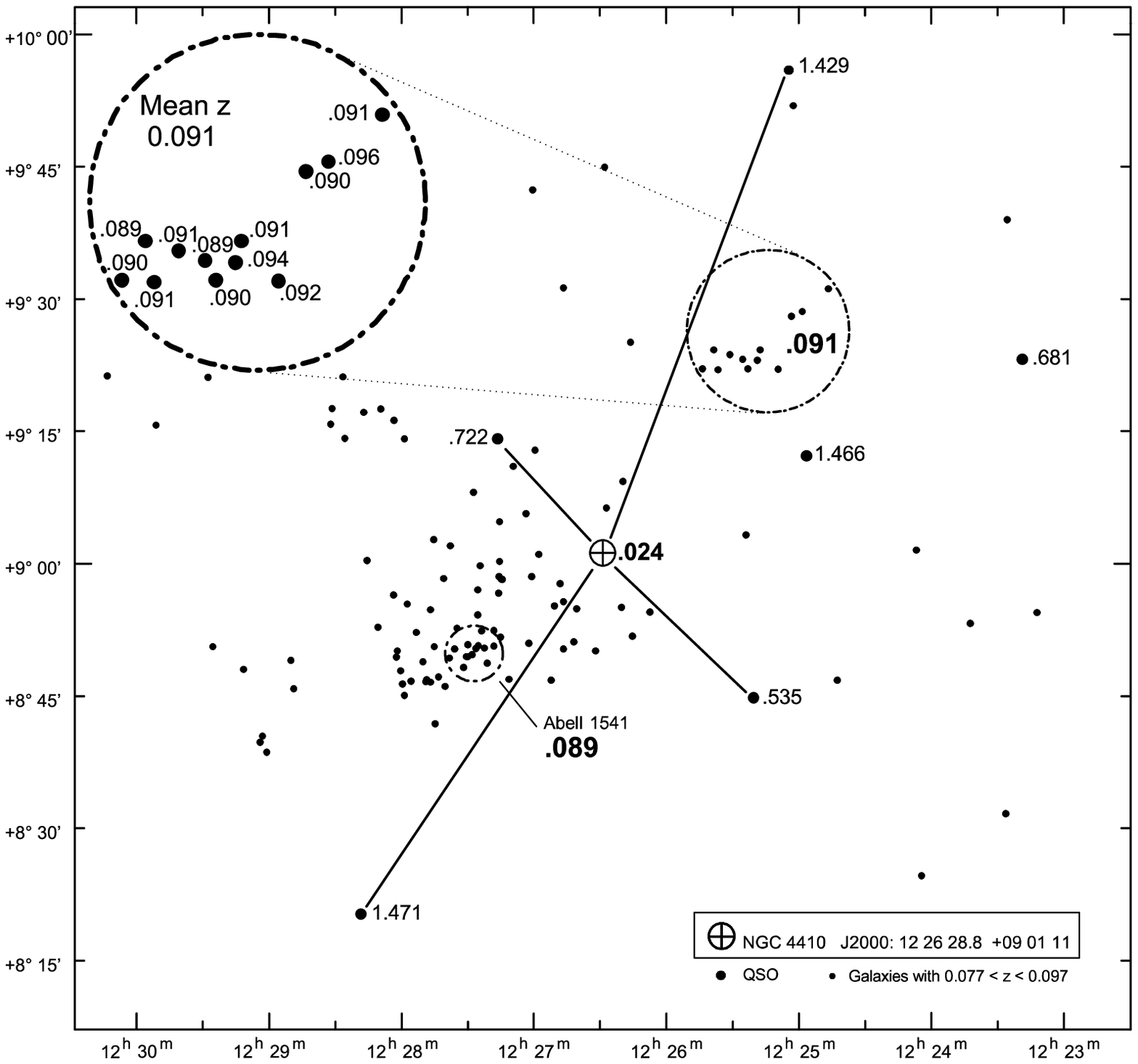}
\caption{All galaxies with $.077 \leq z \leq .097$ within square 120'
on a side around NGC 4410. Circle shows location of Abell 1541 with z
= .089.
\label{fig5}}
\end{figure}

\begin{figure}
\includegraphics[width=11cm]{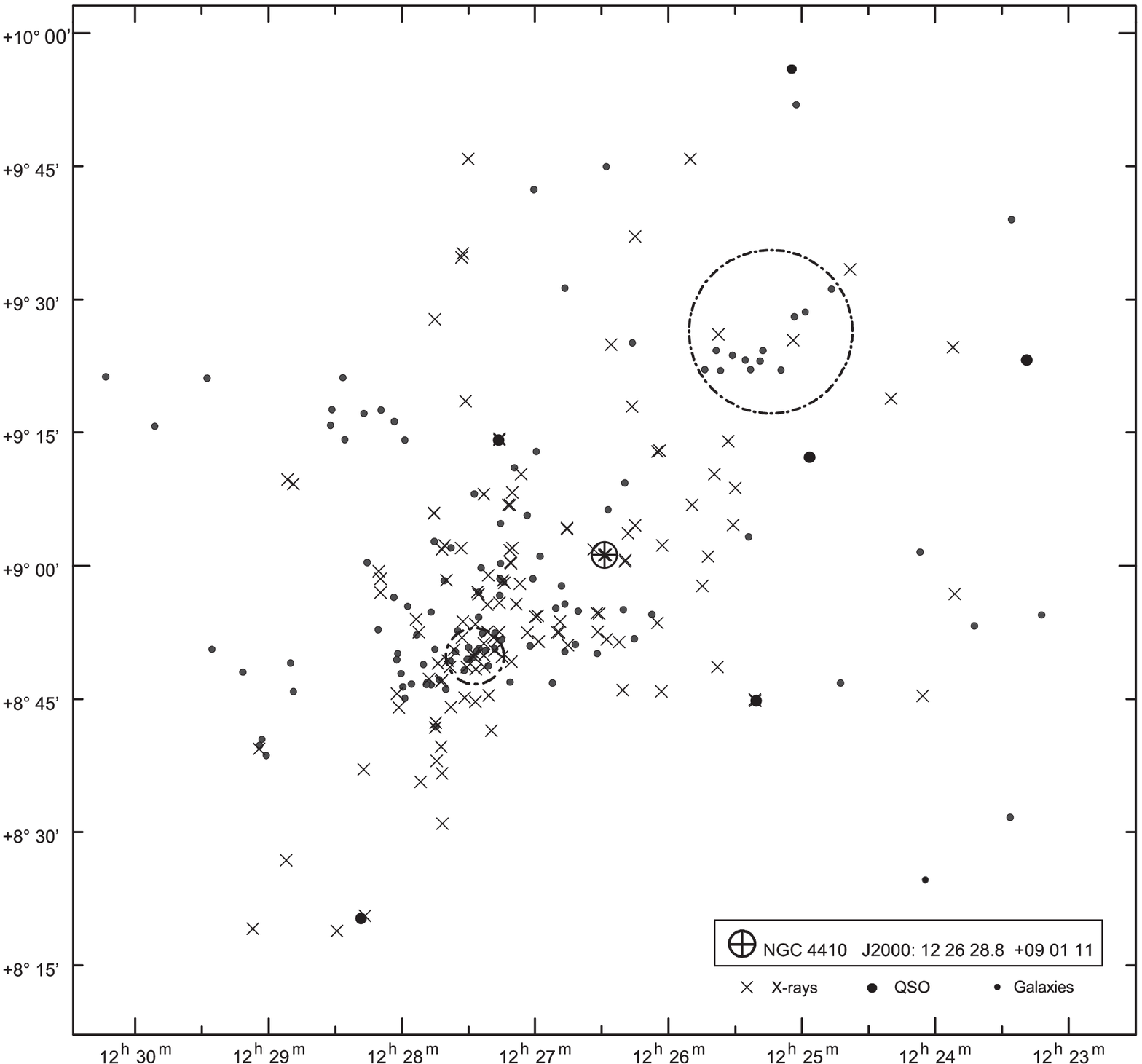}
\caption{The same field as Fig. 5 with X-ray sources added as X
symbols. Data from ROSAT Web Browser.
\label{fig6}}
\end{figure}

\clearpage

\begin{figure}
\includegraphics[width=8cm]{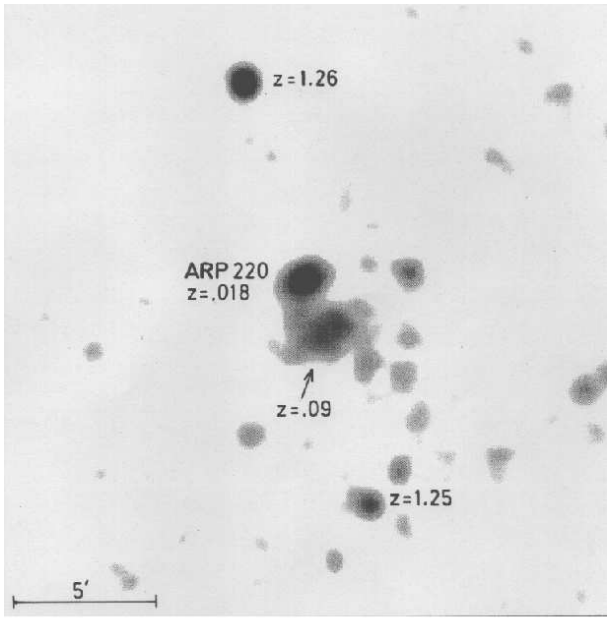}
\caption{An X-ray picture of the ultra luminous, active
galaxy Arp220. Note the emergence of the group of X-ray galaxies on
the line to the z = 1.25 quasar and the many redshift similarities to
the NGC4410 association.
\label{fig7}}
\end{figure}

\medskip

\section{Paper submitted 9 May. Report received 10 May from Astronomy and
Astrophysics} 

We have read attentively your paper "Quasars and Galaxy Clusters Paired Across
NGC 4410" and conclude unfortunately that we cannot accept it on the grounds
that its scientific content is not sufficient to warrant publication in A\&A.

Indeed, the heart of the paper is to investigate all alignment effects around a
precise location on the sky, around a precise galaxy. This is not original,
since your group has claimed alignment for many objects in the past, so there
is nothing new. In addition, it is quite easy to find such alignments in the
sky, given the spatial distribution of galaxies, distributed in a fractal
structure of filaments, great walls, and non-uniform structure, that has now
been even better revealed and precised by large surveys such the SDSS.
  So many remarks of alignment could be noticed like that, and this would be
purely by chance, as can be simulated in numerical simulations of cosmic
filaments.  No new observations are reported here, no new physics is involved
either, and this short note only emphasizes some more numbers and coincidences,
that can appear purely by chance.

 We regret to inform you that we shall be unable to give any further
consideration to this paper.

We are sorry to disappoint you on this occasion.

Yours sincerely,

The Editors

\end{document}